\documentclass[a4paper,11pt]{article}
\pdfoutput=1 % if you are submitting a pdflatex (i.e. if you have
\usepackage{jinstpub} % for details on the use of the package, please
\usepackage{float}
\usepackage{tabularx}
\usepackage{ragged2e} \newcolumntype{L}[1]{>{\raggedright\arraybackslash}p{#1}}
\newcolumntype{C}[1]{>{\centering\arraybackslash}p{#1}}
\newcolumntype{R}[1]{>{\raggedleft\arraybackslash}p{#1}}
\newcolumntype{J}[1]{>{\justifying\arraybackslash}p{#1}}
\usepackage[position=t]{subcaption}
\usepackage{wrapfig}
\newcommand{\Ox}{~O${}_{2}${}}
\newcommand{\Ni}{~N${}_{2}${}}
\newcommand{\Cl}{~Cl${}_{2}${}}
\newcommand{\CaF}{~CaF${}_{2}${}}
\newcommand{\SO}{~SO${}_{2}${}}
\newcommand{\twz}{~cm${}^{-1}${}}
\newcommand{\perc}{\,\%{}}
\newcommand{\mum}{~$\mu$m{}}

\hypersetup{pdfauthor={A. Walter. F. Wilsenack, T. Wolf, F. Duschek},pdftitle={Ultraviolet Raman Spectroscopy for Remote Detection of Chlorine Gas}}
\title{Ultraviolet Raman Spectroscopy for Remote Detection of Chlorine Gas}

\author[a]{A. Walter,} \author[b]{F. Wilsenack,} \author[b]{T. Wolf,} \author[a,1]{F. Duschek\note{Corresponding author.}}

\affiliation[a]{German Aerospace Center, Institute of Technical Physics,\\Im Langen Grund 1, 74239 Hardthausen, Germany} \affiliation[b]{Bundeswehr Research Institute for Protective Technologies and Protection (WIS), 29633 Munster, Germany}

\emailAdd{Frank.Duschek@dlr.de}

\abstract{
As a primary material frequently used in industry, chlorine is relatively easy to obtain and available even in large quantities.
Despite its high toxicity, molecular chlorine is readily available since it is an essential educt in the chemical industry. Over the past decades, numerous accidents involving injured and dead victims have occurred. Furthermore, it was already misused as a warfare agent at the beginning of the last century with still reported attacks.
Early detection, localization, and monitoring of sources and cloud movements are essential for protecting stationary facilities, mobile operations, and the public.
In contrast to most chemical hazardous materials, where it is possible to detect them by vibrational spectroscopic methods (e.\,g., passive hyper-spectral absorption technologies in the infrared), halogens are inactive to infrared absorption. Raman-based technologies rely on changes in the polarizability of the molecule and provide vibrational-spectroscopic access to such diatomic molecules and therefore close the gap in infrared detection capabilities.
Here we present a straightforward approach for a standoff Raman detector in a backscattering configuration. This paper uses a simplified model to discuss optimum excitation wavelengths in achievable detection ranges. We validate the model by spontaneous (vibrational) Raman spectroscopic measurements between 20 and 60~m standoff distance. We also briefly discuss detection performances and technical and physical aspects as prospects of system design.
}


\arxivnumber{4648488} % only if you have one

\proceeding{6$^{\text{th}}$ International Conference on Frontiers of Diagnostic Technologies\\
  \today  \\
  Frascati, Italy}

\begin{document}
\maketitle
\flushbottom

\section{Introduction}
\label{sec:intro}

As a primary material frequently used in industry, chlorine is relatively easy to obtain and available even in large quantities. Chlorine, therefore, poses a risk in the context of industrial accidents (Jordan, 2022), as well as a threat in both terrorist and military scenarios (Iraq and Syria, 2014ff)~\cite{Haber1986, NCSTRT}. Therefore, early detection, localization, and monitoring of sources and cloud movements are essential for protecting stationary facilities, mobile operations, and the public (A cloud at a distance of 1 km traveling at moderate wind speed would take about 2 minutes to reach an observer).
Environmental detection of hazardous gases has been investigated extensively by on-site sensors for monitoring purposes and for standoff solutions reviewed in Ref.~\cite{Hug2015, Gaudio2017, Zhu2018, Jindal2021}.
For detection on-site, a vast variety of technologies have been developed, partly to the level of performing commercial products, some of them also to be integrated into mobile platforms such as unmanned aerial vehicles~\cite{Panne1998, Hug2015}.
Remote sensing (standoff) technologies frequently make use of passive {(hyper-)spectral} approaches~\cite{Zhu2018} or by active illumination of targeted substances with infrared laser (IR) sources~\cite{AugustusWayFountain2009, Li2020}. These IR methods probe vibrational transitions in molecules and benefit from the unique fingerprint structure of the absorption spectra. IR absorption requires a change of the electric dipole moment during the interaction of the electromagnetic wave with the molecule. For highly symmetric molecules such as oxygen, nitrogen, or halogens, it is well known that IR absorption cross sections of their only vibration are by orders of magnitude lower than in less symmetrical molecules~\cite{Gordon2022}.

In higher amounts and with sufficient sunlight, the yellow-green gas can be detected by passive methods~\cite{SURFACEOPTICSCORP2010}. However, active detection technologies are promising solutions at night, especially at lower partial pressures (which are still highly harmful to human beings~\cite{Winder2001}).
Raman spectroscopic research on molecular chlorine and other hazardous gases is well advanced~\cite{Hochenbleicher1971,Edwards1976,Chang1978,Guo2021}. However, few publications on potential standoff applications for diatomic gases (such as hydrogen) are known~\cite{Limery2017,Gallo2021b}. \citeauthor{Eto2022} demonstrate the 20~m standoff detection of \SO { } with a Raman setup in backscattering configuration~\cite{Eto2022} in the deep ultraviolet (UV). The applied excitation wavelength of 217~nm makes a profit of the $1/\lambda^4$-dependency and resonance effects~\cite{Long2002} but would be disadvantageous for detection from larger standoff distances due to absorption of laser and scattered light by atmospheric oxygen~\cite{Sander2009ChemicalKA}.

This work addresses two major points: In the next section, an empirical model is applied to the propagation of light during a (spontaneous) Raman scattering to find a suitable wavelength for the standoff detection of chlorine. The following section demonstrates the remote detection of chlorine spontaneously at a UV excitation wavelength of 266~nm at distances from 20 to 60~m.

\section{Model for standoff Raman detection of chlorine} \label{sec:mat} %\subsection{Model description} %\label{sec:model}
For the first feasibility studies and the design of the experimental geometry and setup, we applied an empirical ray-tracing model that allows for choosing available (and cost-efficient) system components required by Raman detection for chlorine.
\begin{figure}[H]
\centering %
\begin{subfigure}[t]{0.58\textwidth}
\includegraphics[width=1\textwidth]{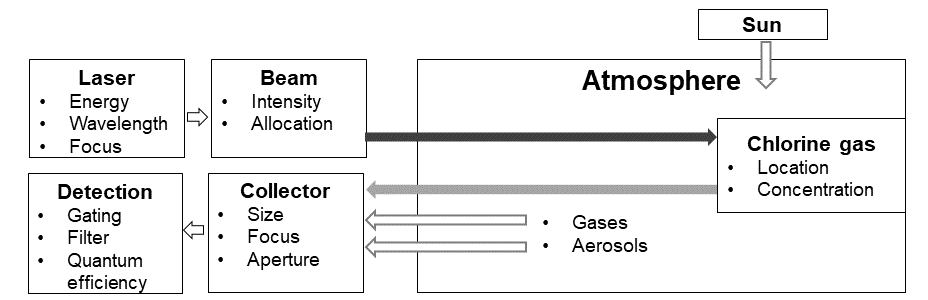}
\caption{Schematical illustration of the model.}
\label{fig:model}
\end{subfigure} \hfill
\begin{subfigure}[t]{0.4\textwidth}
    \includegraphics[width=1\textwidth]{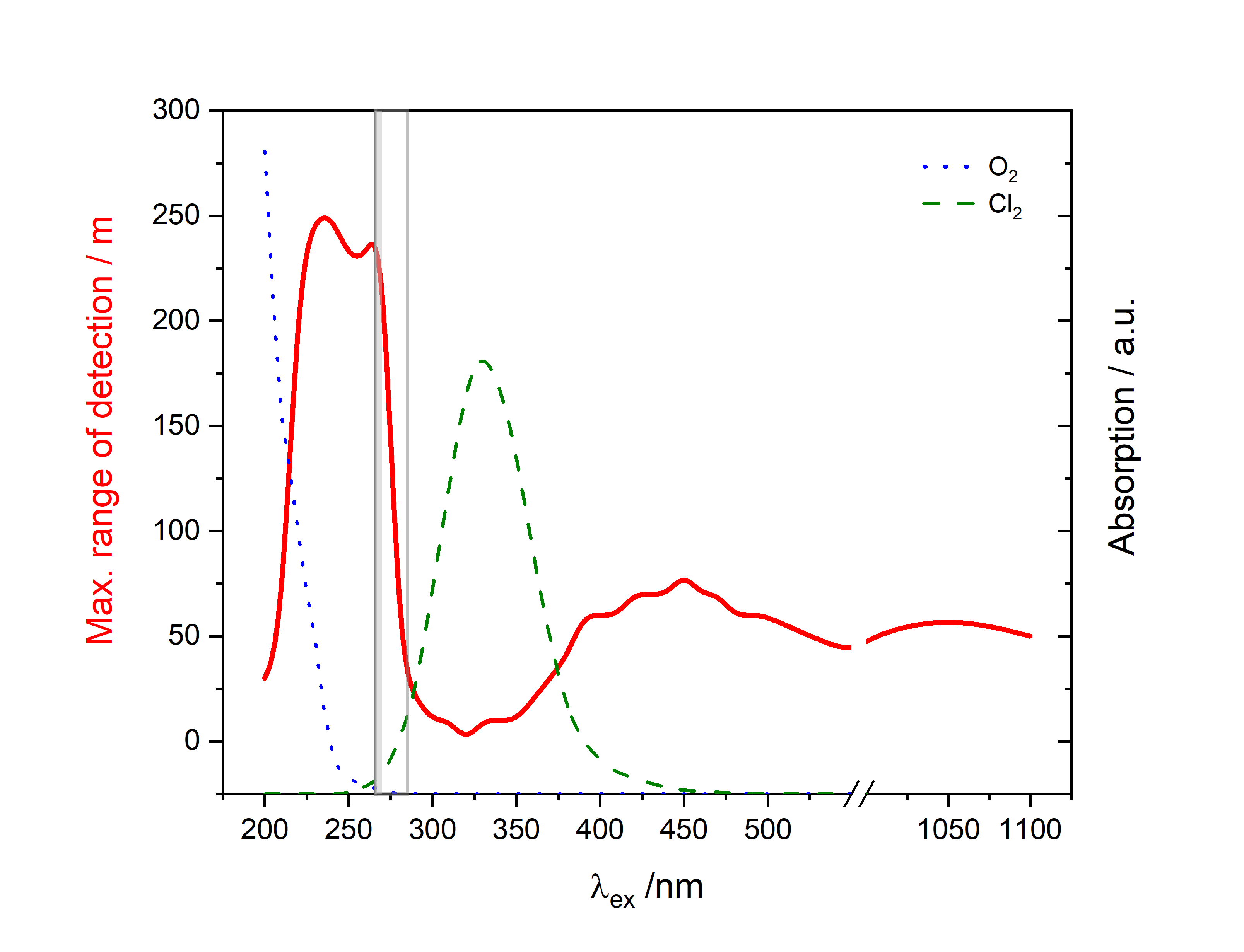}
    \caption{Calculated maximum detection ranges for chlorine gas by Raman spectroscopy as a function of the excitation wavelength.}
    \label{fig:model2.1}
\end{subfigure}
    \caption{Modeling scheme and results.}
\end{figure} The model (see Fig.~\ref{fig:model}) accounts for:
\begin{itemize}
  \setlength\itemsep{0em}
  \item Absorption and scattering effects of the laser and scattered radiation from the detection system to the targeted region and inside a chlorine gas cloud
  \item Expected signals for given laser parameters (pulse energy, repetition rates, wavelength, beam diameter), the detection unit (incl. diameter of collection optics, optical filters, the quantum efficiency of the sensor, time-gating)
  \item Interferences of present light sources (e. g. the sun)
  \item Position and dimension of the chlorine gas cloud source, the chlorine partial pressure, and published absorption and Raman scattering cross-sections.
\end{itemize}
%\subsection{Modeling results}
In this paper, we evaluate the model to find optimized standoff ranges for different Raman excitation wavelengths. The model's parameters (quantum efficiencies, laser pulse energy ...) assume to be market-available and cost-efficient components. For obvious reasons, we don't take Resonance Raman's effects into account.
For the presented results, the model assumes a detection system with a single laser pulse at 100~mJ (unbroadened line width) used for spontaneous Raman excitation of chlorine.
A telescope (400~mm aperture) collects the scattered light. Laser line filters suppress the elastic scattering (optical density OD~6, transmission 95\perc, 1~nm steepness). We detect spectra using a time-gated detector with a quantum efficiency of 20\perc { } in the ultraviolet and 30\perc { } in the visible and near-infrared.
A sphere of 10~m diameter and 0.1\perc { } chlorine partial pressure in the air represents the gas.
We included possible interfering (solar) radiation between 300 and 1000 W/m${}^2$. Detection of the vibrational mode is regarded as successful when > 10 photons are detected at a  signal-to-noise ratio SN~>~2~\cite{Shrivastava2011}.

Fig.~\ref{fig:model2.1} shows the calculated maximum detection distances as a function of excitation wavelengths. Applying Raman spectroscopy at lower excitation wavelengths suffers from atmospheric absorption of the incident and scattered light. On the other hand, at higher excitation wavelengths decreasing Raman scattering (due to $1/\lambda_{ex}^4$-dependency) and increasing interferences from solar radiation compete with the atmospheric transmission. Therefore, we found maximum standoff distances of 280~m in the wavelength region between 240 and 270~nm as a tradeoff.
%\begin{figure}
%\centering
%    \includegraphics[width=0.6\textwidth]{./figs/chlorine-model-distance-vs-wl}
%    \caption{Calculated maximum detection ranges for chlorine gas by Raman spectroscopy as a function of the excitation wavelength.}
%    \label{fig:model2.1}
%\end{figure}
\section{Experimental: spontaneous UV Raman scattering at 266 nm} %\subsection{Experimental Setup} \label{sec:setup}
\begin{figure}[h]
\centering %
\begin{subfigure}[t]{0.45\textwidth}
    \includegraphics[width=\textwidth]{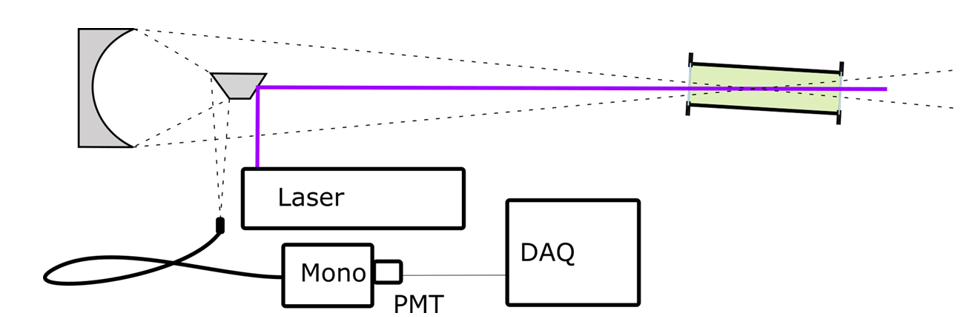}
    \subcaption{Setup for the Raman standoff detection experiment.\label{fig:setup1}}

\end{subfigure}
\hfill
\begin{subfigure}[t]{0.45\textwidth}
\includegraphics[width=\textwidth]{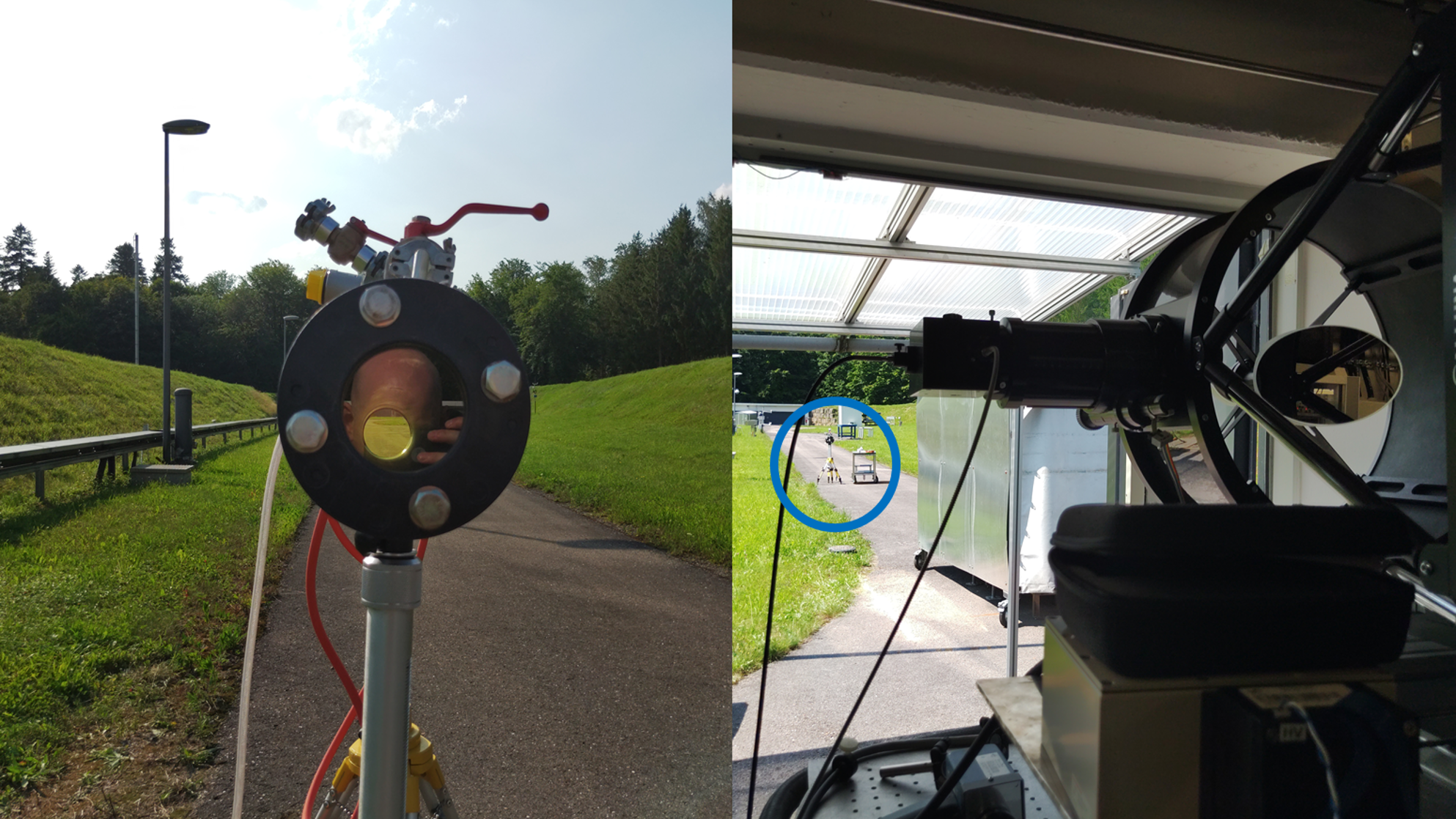}
    \subcaption{Photographes of the sample cell and view along the optical path on the test range.\label{fig:setup2}}

\end{subfigure}
\caption[fig:setup]{Schematic and photographic description of the Raman experiment.} \end{figure}

\begin{figure}[H]
\centering % \begin{center}/\end{center} takes some additional vertical space
\begin{subfigure}[t]{0.45\textwidth}
    \includegraphics[width=\textwidth]{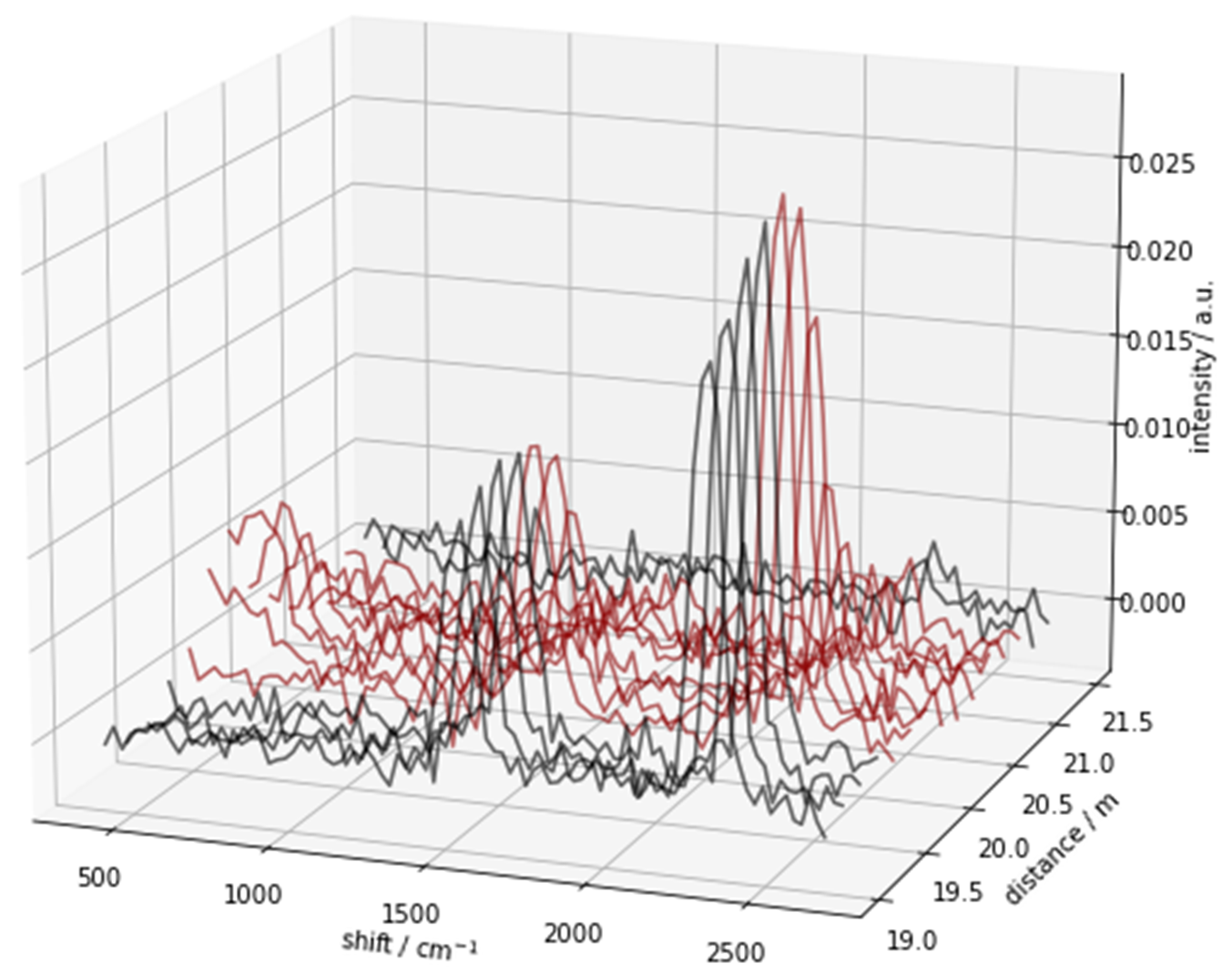}
    \subcaption{Backscattered Raman spectra of chlorine in the air at different standoff distances. Red spectra represent distances in the sample cell.}
    \label{fig:3d}
\end{subfigure}
\hfill
\begin{subfigure}[t]{0.45\textwidth}
\includegraphics[width=\textwidth]{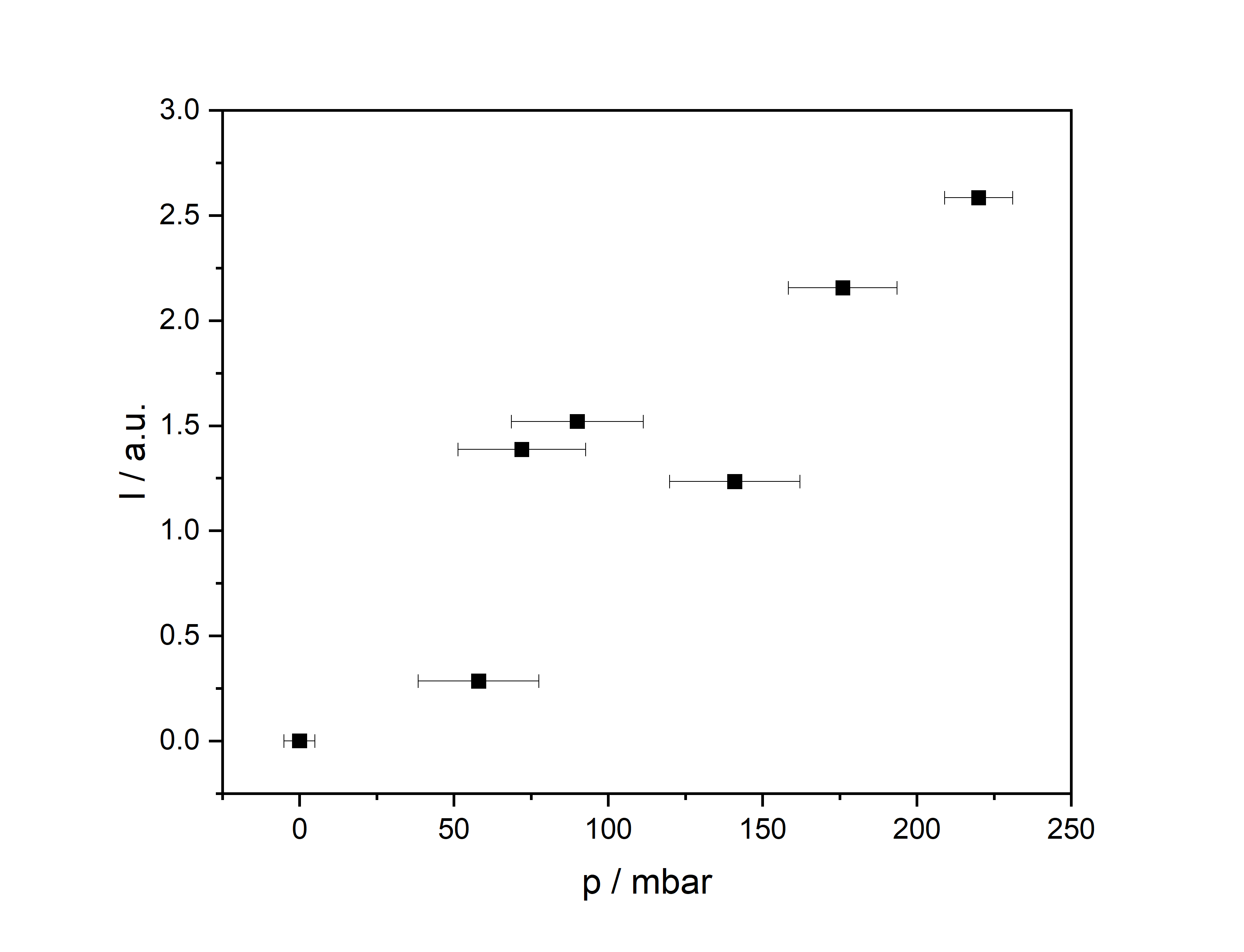}
    \subcaption{
    Maximum intensities of chlorine Raman signals at 20~m for different partial pressures (mbar).}
    \label{fig:res2}
\end{subfigure}
\hfill
\begin{subfigure}[t]{0.45\textwidth}
\includegraphics[width=\textwidth]{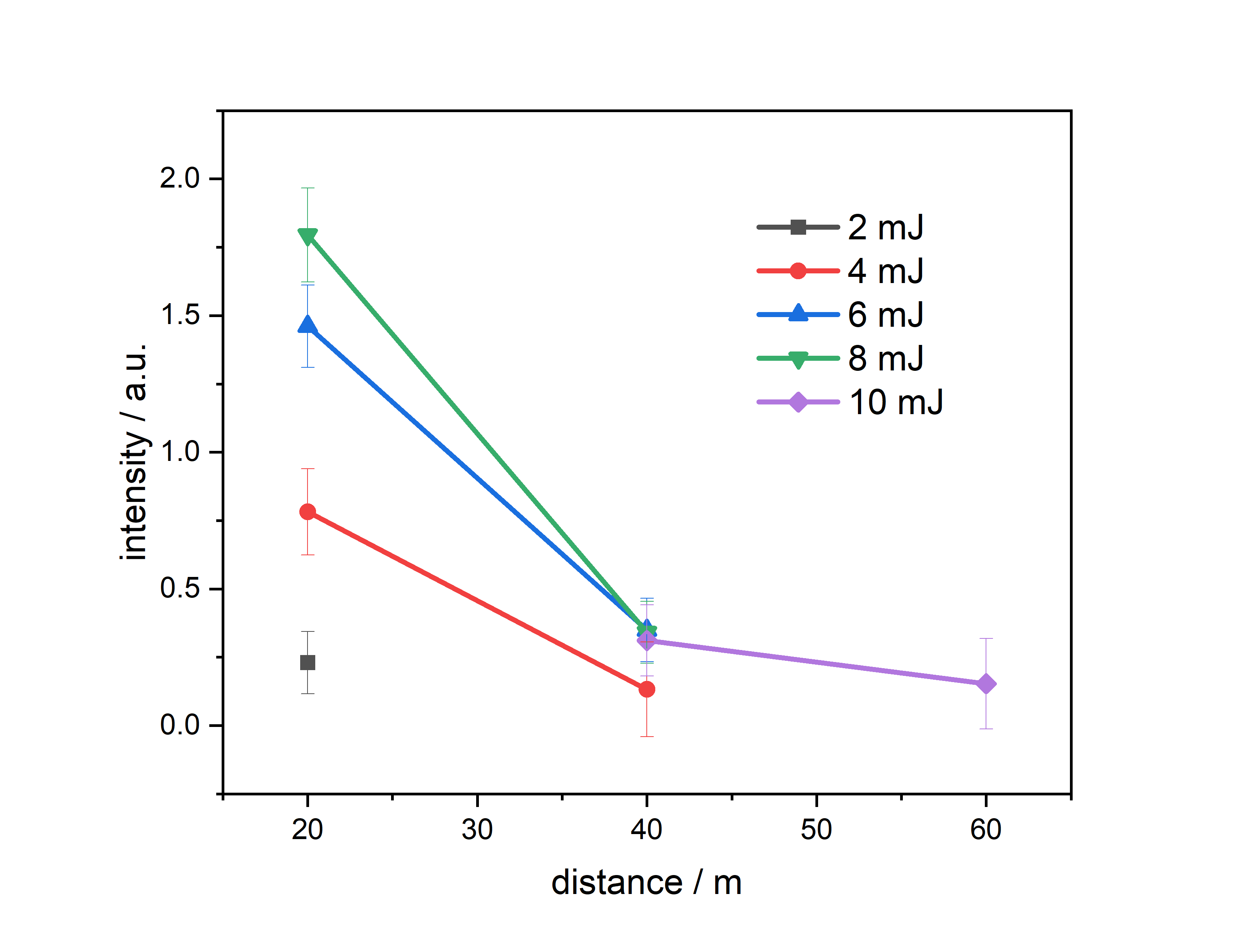}
\caption{
Signal intensities vs. standoff distances from 20, 40, and 60~m.} \label{fig:res3}
\end{subfigure}
\caption[fig:res]{Experimental results for spontaneous Raman scattering of chlorine (200~mbar\Cl, 800~mbar\Ni, 1152 pulses per channel).}
\end{figure}

The experiment (Fig.~\ref{fig:setup1}) applies the 4$^{th}$ harmonic of a diode-pumped Nd:YAG laser (InnoLas SpitLight DPSS EVO IV, 266 nm, repetition rate 100~Hz, pulse length 0.7~ns, pulse energy 10~mJ, line width < 1~nm). The collimated laser beam (10~mm diameter) is aligned co-linear to the central axis of a Newtonian-type telescope and directed onto the sample.
The Raman scattering is collected by the 400~mm Newton telescope and utilizing an optical fiber guided into the spectrometer (Acton SP2500i, 1800 g/mm holographic UV gratings, 200\mum { } slit width) with a spectral resolution of 60\twz. A Hamamatsu photomultiplier tube (Hamamatsu PMT H10720-113
with highspeed amplifier C5594, quantum efficiency  25\perc, response time 0.57~ns) detects the signal. Finally, a storage oscilloscope (Keysight DSOX6004A) records it. An ultrasteep long pass filter blocks scattered laser light at 268.6~nm (Semrock RazorEdge, OD~6, 70\perc { } transmission, 2.1~nm steepness). You should note that the large spectral width of the laser of ~140\twz { } is still sufficient to distinguish between the present atmospheric gases.
For safety reasons, a gas cloud has not been released but is emulated by a sample cell. The cell, a 1000~mm long cylinder with a diameter of 120~mm,  and Raman grade calcium fluoride windows (5~mm thickness, Korth Kristalle, Germany) at the beam entrance and exit~\cite{Gallo2021b}, is tilted by 5\textdegree{} towards the incident beam and is filled with a 20\perc { } mixture of chlorine and dry nitrogen. The \CaF { } windows generated a strong Raman signal at 321\twz, which partly overlapped with the chlorine signal. Therefore, we subtracted, where possible, the isolated window signals. Otherwise, we only attributed signals with a shift >~500\twz { } to chlorine.
For the measurements, each spectrum has typically been accumulated for 10 seconds/channel.
 Backscattered Raman signals are given in Fig.~\ref{fig:3d} for different ranges from 19 to 22 m. Since the toxic chlorine gas was encapsulated into a sample cell placed at a 20~m distance, only \Ox { }and \Ni { } vibrational modes at 1550 and 2300\twz, respectively, can be observed in front of the cell, i.e. up to 19.5~m. At 20~m backscattering, the \Cl { } scattering band appears at 550\twz, slightly overlapped by Raman scattering from \CaF { } windows (321\twz).
While propagating through the sample cell, chlorine gas reduces and absorbs laser and backscattered light.

In Fig.~\ref{fig:res2} we show the dependence of Raman intensities on the chlorine gas concentration. You can observe diminishing returns with higher chlorine partial pressure, which our model can explain by the absorption of the incident laser light and the Raman scattering by chlorine.
%\begin{figure}[H]
%\centering % \begin{center}/\end{center} takes some additional vertical space
%\begin{subfigure}[t]{0.45\textwidth}
%    \includegraphics[width=\textwidth]{./figs/chlorine-3d}
%    \subcaption{Backscattered Raman spectra of chlorine in the air at different standoff distances. Red spectra represent distances in the sample cell.}
%    \label{fig:3d}
%\end{subfigure}
%\hfill
%\begin{subfigure}[t]{0.45\textwidth}
%\includegraphics[width=\textwidth]{./figs/chlorine-model-intensity-vs-p.png}
%    \subcaption{Intensity (a.u.) versus chlorine partial pressure (mbar).}
%    \label{fig:res2}
%\end{subfigure}
%%\hfill
%\caption[fig:res]{Experimental results for spontaneous Raman scattering of chlorine.}
%\end{figure}
%
%\begin{wrapfigure}{r}{0.45\linewidth}
%\includegraphics[width=\linewidth]{./figs/chlorine-model-intensity_vs_distance}
%\caption{Signal intensities vs. standoff distances from 20, 40, and 60 m.} \label{fig:res3}
%\end{wrapfigure}
Fig.~\ref{fig:res3} presents the intensities for the Raman signals at 550\twz { }  for different pulse energies as a function of the standoff distance. The Raman signal shows a decrease according to approximately the $1/r^2$ law.
For the measurements at 20~m, a minimum laser energy of 2~mJ is required for reliable detection of chlorine gas (200~mbar, 1~m 'cloud diameter'). For the larger distances at 40 and 60~m, the required excitation energies are 4~mJ and 10~mJ per pulse, respectively, for the same sample.
At 20~m, the Raman intensity scales linearly with excitation energy.

\section{Summary and conclusion}
This work introduces a simplified model for the design of a standoff chlorine detector by spontaneous Raman spectroscopy. The model considers detection system parameters, atmospheric propagation, and chlorine absorption losses. We should expect this model an enhanced sensitivity at the spectral region between 240 and 270~nm with possible detection ranges of more than 200~m.
We already demonstrated chlorine detection, with a straightforward and not-optimized setup, at standoff distances between 20 and 60~m.

The chosen excitation wavelength of 266~nm (and therefore the wavelength of the Stokes Raman scattering with its maximum at 270~nm) shows promising results but is even at the edge of the best detection range. However, a Raman excitation at a shorter wavelength e.\,g. 248~nm predicts to overcome absorption effects by chlorine and atmospheric gas but will be more demanding in regards of spectral filter. For optimization, we recommend integrating a multichannel sensor with time-gating capabilities into the system~\cite{Koegler2020}. Besides the alarm on the presence of molecular chlorine, information on its amount relative to oxygen and nitrogen can be extracted.
During the Raman experiments, deep UV applications have turned out to be very robust against solar radiation interferences.
Currently, we focus the telescope on a point of interest and collect mainly from the focus.
Switching to gated photon counting technique will allow  to detect at a wider range of distances to be monitored.

%=====================================
% References, variant A: external bibliography %=====================================
\bibliography{articleRXIVE}
\bibliographystyle{IEEEtranN}
\end{document}